\documentclass[12pt]{article}
\usepackage[cp1251]{inputenc}
\usepackage[english,russian]{babel}
\usepackage{cite}

\begin{document}

\title{Quantum Fourier transform and tomographic R\'{e}nyi entropic inequalities}

\author{M. A. Man'ko\thanks{%
P. N. Lebedev Physical Institute, Leninskii Prospect 53, Moscow 119991,
Russia, e-mail: mmanko@sci.lebedev.ru} ,  V. I. Man'ko\thanks{%
P. N. Lebedev Physical Institute, Leninskii Prospect 53, Moscow 119991,
Russia, e-mail: manko@sci.lebedev.ru}}

\date{}

\maketitle

\begin{abstract}
R\'{e}nyi entropy associated with spin tomograms of quantum states is shown
to obey to new inequalities containing the dependence on quantum Fourier transform.
The limiting inequality for the von Neumann entropy of spin quantum states and a new kind
of entropy associated with quantum Fourier transform are obtained. Possible connections
with subadditivity and strong subadditivity conditions for tomographic entropies
and von Neumann entropies are discussed.
\end{abstract}

keywords: uncertainty relations, entropy, quantum tomography, quantum Fourier transform.

\section{Introduction}

In previous work \cite{Rui-JRLR}, the probability-operator-symbol framework for
quantum information was constructed. In this approach, quantum-information ingredients
like qubits, qudits, and the operators providing the description and connection of the
qudit states are given in the form of functions called the operator symbols.
For the operator symbols, the product rule called the star-product is determined by using
an integral nonlocal kernel~\cite{Beppe-OlgaJPA}. For the both pure and mixed qudit states,
the operator symbols of their density operators are standard probability-distribution
functions. Since the qudit states for multipartite systems are described by standard
probability distributions, all the characteristics of the distributions including
Shannon entropy~\cite{shannon} and R\'{e}nyi entropy~\cite{renyi} can be used to introduce
in quantum information~\cite{Rui-JRLR} the operator-symbol entropies like, e.g.,
operator-symbol R\'{e}nyi entropies, operator-symbol relative $q$-entropy. Since the
Shannon entropy is the limiting case of the R\'{e}nyi entropy, the corresponding analogs of
the operator-symbol Shannon entropy and the properties of this entropy can be
obtained in quantum information within the introduced operator-symbol framework.

In quantum mechanics and quantum information, there is a fundamental feature
distinguishing their quantum nature from their classical counterparts, that is the
uncertainty relations. The uncertainty relations by Heisenberg \cite{heis27} and
by Schr\"odinger \cite{schrod30} and Robertson \cite{rob30,rob34} written for conjugate
variables like positions and momenta were also accompanied by the so-called entropic
uncertainty relations. The entropic uncertainty relations for the continuous variables
were written in the form of inequalities for Shannon entropy associated with the position
and momentum probability densities in \cite{hir,Bal2,183}.

Within the operator symbol framework, in quantum information the subadditivity and strong
subadditivity conditions were obtained for probability distribution describing qudit
quantum states (called spin tomograms)~\cite{Rui-JRLR}. Also some relations of the
inequalities with subadditivity and strong subadditivity conditions for the von Neumann
entropy were clarified. In fact, within the operator symbol framework, the tomographic map
of unitary group $U(n)$ onto simplex was given and, in view of this map, the notion of
Shannon entropy, R\'{e}nyi entropy, and other entropies were introduced for unitary group.

The entropic uncertainty relations for finite-dimensional quantum systems were obtained
in the form of inequalities for Shannon entropies associated with probability distributions
related to measuring noncommuting observables in \cite{uff,Raj,RuiZ,Balles,Azar}.
The uncertainty relations for R\'{e}nyi entropy for the position and momentum distributions
and for finite-dimensional systems with measuring observables related by quantum Fourier
transform were obtained in \cite{BiruPRA}. There exist for bipartite and tripartite systems
the known inequalities for the von Neumann entropy called subadditivity and strong subadditivity
conditions \cite{RusLieb,Rus}.

The aim of this work is to review the results of \cite{Rui-JRLR} and extend the study
of entropic inequalities like subadditivity and strong subadditivity conditions obtained in
the previous work on the probabilistic operator symbol framework in quantum information and
to find new entropic inequalities for spin tomograms which are analogs of entropic inequalities
discussed in \cite{uff,Raj,RuiZ,Balles,Azar,BiruPRA}. For continuous variables, the entropic
inequalities for quantum symplectic tomogram were discussed in \cite{Ren1,VinoSem,Ren2,Ren3,Ren4,Ren5}.
The essential aspect of the new inequalities is that they are closely related to properties
of quantum Fourier transform discussed in \cite{schwinger} and used in quantum information
(see, for example, \cite{vourdas,Shor-alg,weinst}). The quantum Fourier transform as an
important ingredient plays a key role in quantum computing, quantum information processing.\footnote{
This work was initiated by illuminating discussions with Rui Vilela Mendes and we thank him
for his help in obtaining the results presented here.}

We will get entropic inequalities which provide some constrains for unitary spin tomograms
connecting them with $\ln N$, where $N$ is dimension of Hilbert space.
The number $\ln N$ has appeared in entropic inequalities in earlier works~\cite{uff,Raj,RuiZ,Azar}
but a new element of the present study is that $\ln N$ is directly associated with
tomographic-probability distribution determining the qudit state in the probability operator
symbol framework in quantum information.

\section{Spin tomograms}

Given an $N$-dimensional Hilbert space of spin system. One can interpret the Hilbert
space either as the state space for one particle with spin $j=(N-1)/2$ or in the case
of product representation of number $N=n_1n_2\ldots n_M$ as the space of multipartite
spin system with $j_1=(n_1-1)/2,j_2=(n_2-1)/2,\ldots,j_M=(n_M-1)/2$.
The $N$$\times$$N$ density matrix $\rho$ of the quantum state can be represented by the
unitary tomogram of the spin state \cite{SudPLA}.

In the case of spin state with $j=(N-1)/2$, the tomogram is defined by the relation
\begin{equation}\label{1}
w(m,u)=\langle m\mid u^\dagger\rho u\mid m\rangle,
\end{equation}
where $\rho$ is the density matrix, $u$ is $N$$\times$$N$ unitary matrix and semi-integers
$m=-j,-j+1,\ldots,j$ are values of spin projection on the $z$ axis. The tomogram $w(m,u)$
is nonnegative probability distribution function of random spin variable satisfying the
normalization condition
\begin{equation}\label{2}
\sum_{m=-j}^jw(m,u)=1
\end{equation}
and the equality
\begin{equation}\label{3}
\int w(m,u)\,du=1,
\end{equation}
where $du$ is Haar measure on unitary group with normalization
\begin{equation}\label{4}
\int du=1.
\end{equation}
The important property of tomogram (\ref{1}) is that its connection with density matrix
$\rho$ is bijective, i.e., $\rho\leftrightarrow w(m,u)$ \cite{Rui-JRLR}. This means
that the quantum state is given if the tomogram is known \cite{DodPhysLett,OlgaJETP}.

\section{Quantum Fourier transform}

The symmetric unitary $N$$\times$$N$ matrix $F$ with matrix elements
$F_{jk}=\frac{1}{\sqrt N}\,\exp\left(\frac{2\pi i}{N}jk\right)$
$(j,k=0,1,\ldots,N-1)$, which are characters of irreducible representation of
cyclic group $C_N$, can be used to provide invertible map of normalized complex
vector $\vec a$ with components $a_k$ onto complex vectors $\vec a^{(f)}$ with
components $a_k^{(f)}$ as follows:
\begin{equation}\label{5}
a_k^{(f)}=\sum_{j=0}^{N-1}F_{kj}a_j,\qquad a_k=\sum_{j=0}^{N-1}(F^\dagger)_{kj}
a_j^{(f)}.
\end{equation}
The matrix $F$ with matrix elements $F_{kj}$ satisfies the equality
\begin{equation}\label{6}
F^N=1.
\end{equation}
The map (\ref{5}) is called the quantum Fourier transform for cyclic group $C_N$.

If one uses the usual labels for spin projection $m=-j,-j+1,\ldots,j$, the operator
$\hat F$ for quantum Fourier transform can be defined as
\begin{equation}\label{7}
\hat F\mid m\rangle=\sum_{m'=-j}^jF_{m'm}\mid m'\rangle,
\end{equation}
where the symmetric matrix
\begin{equation}\label{8}
F_{m'm}=\langle m'\mid\hat F\mid m\rangle
\end{equation}
has the form
\begin{equation}\label{9}
F_{m'm}=\frac{1}{\sqrt N}\left(\begin{array}{ccccc}
1 & 1 & 1 & \cdots & 1\\
1 & a & a^2 & \cdots & a^{N-1}\\
1 & a^2 & a^4 & \cdots & a^{N-2}\\
\cdots & \cdots & \cdots & \cdots & \cdots\\
1 & a^{N-1} & a^{N-2} & \cdots & a
\end{array}\right),
\qquad a=\exp\left(\frac{2\pi i}{N}\right).
\end{equation}
Thus the unitary operator of quantum Fourier transform reads
\begin{equation}\label{10}
\hat F=\sum_{m=-j}^j\sum_{m'=-j}^jF_{m'm}\mid m'\rangle\langle m\mid.
\end{equation}
In view of (\ref{6}), one has $(\hat F)^N=\hat 1$,
where $\hat 1$ is the identity operator.

\section{Shannon and R\'{e}nyi tomographic entropies}

Following standard definitions of probability theory, one can introduce
Shannon \cite{shannon} tomographic entropy \cite{Olga2,Rui-JRLR} and
R\'{e}nyi \cite{renyi} tomographic entropy \cite{Rui-JRLR}.
The Shannon tomographic entropy (operator-symbol Shannon entropy) is
the function on unitary group
\begin{equation}\label{12}
H_{u}=-\sum_{m=-j}^jw(m,u)\ln w(m,u).
\end{equation}
The R\'{e}nyi tomographic entropy (operator symbol R\'{e}nyi entropy)
is also the function on the unitary group and it depends on extra parameter
\begin{equation}\label{13}
R_{u}=\frac{1}{1-q}\ln \left(\sum_{m=-j}^j\left(w(\left(m,u\right)\right)^q
\right). \end{equation}
Likewise for two spin tomograms $w_1(m,u)$ and
$w_2(m,u)$, we define the operator symbol relative $q$-entropy
\begin{equation}\label{14}
H_{q}(w_{1}(u)|w_{2}(u))=-\sum_{m=-j}^jw_1(m,u)\ln _{q}
\frac{w_{2}(m,u)}{w_{1}(m,u)}
\end{equation}
with
\begin{equation}\label{15}
\ln_qx=\frac{x^{1-q}-1}{1-q}\,,\quad x>0,\quad q>0,\quad
\ln_{q\rightarrow 1}x=\ln x.
\end{equation}
The relative tomographic $q$-entropy is nonnegative.

For $q\to 1$, $R_u\to H_u$ and the operator symbol relative $q$-entropy becomes
operator symbol relative entropy
\begin{equation}\label{16}
H(w_{1}(u)|w_{2}(u))=-\sum_{m=-j}^jw_1(m,u)\ln\frac{w_{2}(m,u)}{w_{1}(m,u)}. \end{equation}
As was shown in \cite{Rui-JRLR}, the minimum over unitary group of
the operator symbol R\'{e}nyi entropy is equal to quantum R\'{e}nyi tomographic entropy
\begin{equation}\label{17}
\min\,R_{u}=\frac{1}{1-q}\ln \mbox{Tr}\,\rho^q.
\end{equation}
Also the relative entropy (\ref{14}) is nonnegative function
for any admissible deformation parameter $q$.

The minimum of entropy $H_u$ given by (\ref{12}) over the unitary group is equal to
von Neumann entropy \cite{Olga2,Rui-JRLR}, i.e.,
\begin{equation}\label{18}
\min\,H_{u}=-\mbox{Tr}\,\rho\,\ln\rho.
\end{equation}

\section{Shannon entropic inequalities in measuring\\ noncommutative observables}

In this section, we review known entropic inequalities \cite{uff,Raj,RuiZ,Balles,Azar}
which appear in the problem of measuring two observables $\hat A$ and $\hat B$
in finite Hilbert space.

Let the spectral decompositions of Hermitian operators $\hat A$ and $\hat B$ read
\begin{equation}\label{UU1}
\hat A=\sum_kA_k\mid a_k\rangle\langle a_k\mid,
\quad\hat B=\sum_kB_k\mid b_k\rangle\langle b_k\mid,
\quad k=1,\ldots,N,
\end{equation}
where $A_k$ and $B_k$ are eigenvalues of the observables and $\mid a_k\rangle$ and
$\mid b_k\rangle$ are their orthonormal systems of eigenvectors.

For pure state $\mid\psi\rangle$, one has two probability distributions
\begin{equation}\label{UU2}
p_k=|\langle a_k\mid\psi\rangle|^2,\quad q_k=|\langle b_k\mid\psi\rangle|^2.
\end{equation}
The corresponding Shannon entropies connected with these two distributions read
\begin{equation}\label{UU3}
H_{p}=-\sum_kp_k\,\ln p_k
\end{equation}
and
\begin{equation}\label{UU4}
H_{q}=-\sum_kq_k\,\ln q_k.
\end{equation}
They satisfy the inequality found in \cite{deutsch}
\begin{equation}\label{UU5}
H_p+H_{q}\geq -2\ln\frac{1}{2}(1+c),
\end{equation}
where the bound $c$ is determined by maximum values of scalar product modulus
\begin{equation}\label{UU6}
c=\max_{j,k}|\langle a_j\mid b_k\rangle|.
\end{equation}
In \cite{kraus} the inequality was conjectured to be improved
\begin{equation}\label{UU7}
H_p+H_{q}\geq -2\ln c
\end{equation}
and in \cite{uff} it was proved.

For the case of observables $\hat A$ and $\hat B$ with eigenvectors providing
mutually unbised bases $\mid a_k\rangle,\mid b_k\rangle$ (see \cite{schwinger,RuiZ}), i.e.,
\begin{equation}\label{UU8}
\max|\langle a_i\mid b_j\rangle|=\frac{1}{\sqrt N}\,,
\end{equation}
inequality (\ref{UU7}) reads \cite{RuiZ}
\begin{equation}\label{UU9}
H_p+H_{q}\geq\ln N.
\end{equation}
Thus the dimensionality of Hilbert space $N$ appears in the entropic inequality.

The problem of mutually unbiased bases is related to geometry of finite Hilbert
spaces \cite{woot1,woot2}. It was widely discussed in connection with constructing
the Wigner function for finite Hilbert space and quantum cryptography (see, for example,
\cite{kibler,planat,Klimov,vourdas,paz}).

The entropic inequalities for Shannon entropy can be also obtained in studying the problem
of measuring several noncommutative observables with orthonormal sets of eigenvectors which
satisfy the condition (\ref{UU8}) (see \cite{RuiZ,Balles}). In \cite{Raj} the entropic
inequalities for Tsallis entropy related to continuous variables were obtained on the base
of Sobolev inequalities while in \cite{BiruPRA} the analogous entropic uncertainty relations
for R\'{e}nyi entropy both for finite Hilbert space and for continuous variables were presented.

\section{Known inequalities for bipartite and tripartite systems}

The operator symbol entropies satisfy some known inequalities found in \cite{Rui-JRLR}.
For example, if the spin system is bipartite, i.e., one has spin $j_1$ and $j_2$, the
basis in tensor product space reads
\begin{equation}\label{19}
\mid m_1m_2\rangle=\mid m_1\rangle\mid m_2\rangle.
\end{equation}
In this case, the tomogram is the joint-probability distribution of two random
spin projections $m_1=-j_1,-j_1+1,\ldots,j_1$ and $m_2=-j_2,-j_2+1,\ldots,j_2$
depending on $(2j_1+1)(2j_2+1)$$\times$$(2j_1+1)(2j_2+1)$ unitary matrix $u$.
The tomogram reads
\begin{equation}\label{20}
w(m_1,m_2,u)=\langle m_1m_2\mid u^\dagger\rho(1,2)u\mid m_1m_2\rangle,
\end{equation}
where $\rho(1,2)$ is the density matrix of bipartite system with matrix elements
\begin{equation}\label{21}
\rho(1,2)_{m_1m_2,m'_1m'_2}=\langle m_1m_2\mid\rho(1,2)\mid m'_1m'_2\rangle.
\end{equation}

For the tomogram, one can introduce the Shannon entropy $H_{12}(u)$ as
\begin{equation}\label{22}
H_{12}(u)=-\sum_{m_1=-j_1}^{j_1}\sum_{m_2=-j_2}^{j_2}
w(m_1,m_2,u)\ln w(m_1,m_2,u)
\end{equation}
and the entropy satisfies the subadditivity condition for all elements of the unitary
group
\begin{equation}\label{23}
H_{12}(u)\leq H_{1}(u)+H_{2}(u),
\end{equation}
where $H_{1}(u)$ and $H_{2}(u)$ are Shannon entropies associated with subsystem
tomograms
\begin{equation}\label{24}
w_1(m_1,u)=\sum_{m_2=-j_2}^{j_2}w(m_1,m_2,u)
\end{equation}
and
\begin{equation}\label{25}
w_2(m_2,u)=\sum_{m_1=-j_1}^{j_1}w(m_1,m_2,u)
\end{equation}
as follows:
\begin{equation}\label{26}
H_k(u)=-\sum_{m_k=-j_k}^{j_k}w_k(m_k,u)\,\ln w_k(m_k,u),
\qquad k=1,2.
\end{equation}
From this inequality, in view of the relation between the von Neumann and operator
symbol entropies, follows the known inequality \cite{Rui-JRLR}, which is subadditivity condition
for corresponding von Neumann entropy for bipartite system
\begin{equation}\label{27}
S_{12}\leq S_{1}+S_{2},
\end{equation}
where
\begin{equation}\label{28}
S_{k}=-\mbox{Tr}\,\rho_k\,\ln\rho_k, \qquad k=1,2
\end{equation}
and
\begin{equation}\label{29}
\rho_{1}=-\mbox{Tr}_2\,\rho(1,2), \qquad \rho_{2}=-\mbox{Tr}_1\,
\rho(1,2).
\end{equation}

For tripartite spin system with spins $j_1$, $j_2$, $j_3$ and density matrix $\rho(1,2,3)$,
the spin tomogram reads
\begin{equation}\label{30}
w(m_{1},m_{2},m_3,u)=\langle m_{1}m_{2}m_3\mid u^{\dagger }
\rho(1,2,3)u\mid m_{1}m_{2}m_3\rangle.
\end{equation}
One associates with this tomogram the Shannon entropy $H_{123}(u)$.
This entropy satisfies inequality, which is the strong subadditivity condition
on the unitary group. It reads \cite{Rui-JRLR}
\begin{equation}\label{31}
H_{123}(u)+H_2(u)\leq H_{12}(u)+H_{23}(u),
\end{equation}
where
\begin{equation}\label{32}
H_{123}(u)=-\sum_{m_1=-j_1}^{j_1}\sum_{m_2=-j_2}^{j_2}\sum_{m_3=-j_3}^{j_3}
w(m_1,m_2,m_3,u)\ln w(m_1,m_2,m_3,u)
\end{equation} and entropies $H_{12}(u)$, $H_{23}(u)$, and $H_{2}(u)$ are defined by
means of projected tomograms
\begin{eqnarray}\label{32a}
w_{12}(m_1,m_2,u)=\sum_{m_3=-j_3}^{j_3}w(m_1,m_2,m_3,u), \label{33}\\
w_{23}(m_2,m_3,u)=\sum_{m_1=-j_1}^{j_1}w(m_1,m_2,m_3,u), \label{34}\\
w_{2}(m_2,u)=\sum_{m_1=-j_1}^{j_1}w_{12}(m_1,m_2,u).
\label{35}\end{eqnarray}
New inequality (\ref{31}) does not contradict the known strong subadditivity condition
for von Neumann entropy \cite{RusLieb,Rus}
\begin{equation}\label{36}
S_{123}+S_2\leq S_{12}+S_{23},\end{equation}
where
\begin{equation}\label{37}
S_{123}=-\mbox{Tr}\,\rho_{123}\,\ln\rho_{123},
\end{equation}
and other entropies are von Neumann entropies for reduced density matrices
$\rho(1,2)=\mbox{Tr}_3\rho(1,2,3)$ and $\rho(2,3)=\mbox{Tr}_1\rho(1,2,3)$.

Inequalities (\ref{23}) and (\ref{31}) are new inequalities for composite quantum
finite-dimensional systems obtained in \cite{Rui-JRLR}.

\section{New inequalities for R\'{e}nyi operator symbol entropies}

In this section, we continue the study of tomographic entropies along the line
of our previous work \cite{Rui-JRLR} and derive new inequalities for spin tomographic
entropies related to quantum Fourier transform. For continuous conjugate variables
(position and momentum), the inequalities for R\'{e}nyi entropy associated with
probability densities in position and momentum were obtained in \cite{BiruPRA}.
These inequalities were used to obtain new integral inequalities for symplectic
and optical tomograms in \cite{Ren1,Ren2,Ren3,Ren4,Ren5}. In \cite{BiruPRA} for
$N$-dimensional Hilbert space the analog of uncertainty relation for the R\'{e}nyi
entropies was given in the form
\begin{equation}\label{CC1}
\frac{1}{1-\alpha}\,\ln\left(\sum_{k=1}^{N}\widetilde p_k^\alpha\right)+
\frac{1}{1-\beta}\,\ln\left(\sum_{l=1}^{N}p_l^\beta\right)\geq\ln N,
\end{equation}
where
\begin{equation}\label{CC2}
\widetilde p_k=\mid\widetilde a_k\mid^2,\qquad p_l=\mid a_l\mid^2,
\qquad\frac{1}{\alpha}+\frac{1}{\beta}=2,
\end{equation}
and the complex numbers $\widetilde a_k$ and $a_l$ are connected by the quantum
Fourier transform
\begin{equation}\label{CC3}
\widetilde a_k=\frac{1}{\sqrt N}\sum_{l=1}^{N}\exp\left(\frac{2\pi ikl}{N}\right)a_l.
\end{equation}
Below we use the inequalities to obtain new inequalities for Shannon and R\'{e}nyi
entropies associated with unitary spin tomograms. The spin tomogram of a state with
density operator $\rho$ can be considered as a column probability vector on unitary
group with the components $w_m(u)$. Then we can introduce another $N$-vector with
components $p_m(u)=\sqrt{w_m(u)}$. Applying inequality (\ref{CC1}) to these vectors
and the notation
\begin{equation}\label{KK1}
\left|\sum_{m'=-j'}^jF_{mm'}\sqrt{w(m',u)}\right|=\sqrt{w_F(m,u)},
\end{equation}
where $F_{mm'}$ is given by (\ref{9}) and $w_F(m,u)$ is the probability distribution,
we obtain inequality
\begin{equation}\label{KK2}
\frac{1}{1-\alpha}\,\ln\left(\sum_{m=-j}^{j}w(m,u)^\alpha\right)+
\frac{1}{1-\beta}\,\ln\left(\sum_{m=-j}^{j}w_F(m,u)^\beta\right)\geq\ln N.
\end{equation}
Also using for pure state $\mid\psi\rangle$ the definition of spin tomogram, we obtain
another similar inequality
\begin{equation}\label{CC5}
\frac{1}{1-\alpha}\,\ln\left(\sum_{m=-j}^{j}w(m,u)^\alpha\right)+
\frac{1}{1-\beta}\,\ln\left(\sum_{m=-j}^{j}w(m,Fu)^\beta\right)\geq\ln N,
\end{equation}
where $F$ is quantum Fourier transform matrix. We can conjecture that the above inequality
(\ref{CC5}) is valued also for mixed state.

Thus one has for R\'{e}nyi entropy (\ref{13}) the inequality for each unitary matrix
\begin{equation}\label{CC5a}
R_\alpha(u)+R_\beta(Fu)\geq\ln N.\end{equation}
Thus the unitary spin tomogram of the particle with spin $j$ for the state with
$N$$\times$$N$ density matrix $\rho$, where $N=2j+1$, must satisfy inequality
(\ref{CC5}). In the limit $\alpha \to 1$, $\beta \to 1$, one gets inequalities
for Shannon entropy of the spin state
\begin{equation}\label{CC6}
H(u)+H(Fu)\geq \ln N.\end{equation}
Another inequality reads
\begin{equation}\label{CC6a}
H(u)+H_F(u)\geq \ln N,\end{equation}
where $H_F(u)$ is the Shannon entropy associated with the probability distribution $w_F(m,u)$.

For the minimum value of the Shannon entropy realized for unitary matrix $u_0$,
one has the von Neumann entropy
\begin{equation}\label{CC7}
H(u_0)=S_{\rm vN}.
\end{equation}
Inequality (\ref{CC6}) written for $u_0$
\begin{equation}\label{CC7a}
H(u_0)+H(Fu_0)\geq \ln N
\end{equation}
provides the inequality for the von Neumann entropy
\begin{equation}\label{CC8}
S_{\rm vN}+S(Fu_0)\geq \ln N,
\end{equation}
where $S(Fu_0)$ is a new entropy. It has the following physical meaning.
If the density operator $\hat\rho$ of the quantum state of spin is given in the form
of spectral decomposition
\begin{equation}\label{CC9}
\hat\rho=\sum_{q=-j}^{j}\lambda_q\mid q\rangle\langle q\mid,
\end{equation}
one can identify the eigenstate $\mid q\rangle$ of the density operator $\hat\rho$ with
``position'' state. In the approach with mutually unbiased bases
and Wigner function for finite Hilbert space \cite{schwinger,vourdas,kibler,planat,
Klimov,paz,Rosu}, the states
\begin{equation}\label{CC10}
\mid p\rangle=\hat F\mid q\rangle,
\end{equation}
where $\hat F$ is the Fourier transform operator, are interpreted as ``momentum''
eigenstates. The matrix elements
\begin{equation}\label{CC11}
\langle p\mid\hat F\mid q\rangle=F_{pq}
\end{equation}
provide the matrix $F$ which coincides with the Fourier transform matrix.
Thus we have the interpretation of the new inequality in the same manner as it
was done in the case of continuous variables.
The new entropy $S(Fu_0)$ in (\ref{CC8}) is the Shannon entropy for ``momentum''
distribution, if we identify the standard von Neumann entropy with Shannon entropy
for ``position'' distribution.

Let us consider the example of qubit state with density matrix
\begin{equation}\label{CC12}
\rho=\left(\begin{array}{cc}
1 & 0\\
0& 0\end{array}\right).
\end{equation}
Position operator $\hat q$ is $\sigma_z$ matrix and momentum operator $\hat p$ is
$\sigma_x$ matrix. Two position eigenvectors $\mid q\rangle$ are
$\left(\begin{array}{c}1\\0\end{array}\right)$ and
$\left(\begin{array}{c}0\\1\end{array}\right)$ and two momentum
eigenvectors $\mid p\rangle$ are $\frac{1}{\sqrt 2}\left(\begin{array}{c}
1\\1\end{array}\right)$ and $\frac{1}{\sqrt 2}\left(\begin{array}{c}
1\\-1\end{array}\right)$. The matrix $F$ reads
\begin{equation}\label{CC13}F=\frac{1}{\sqrt 2}\left(\begin{array}{cc}
1 & 1\\1 & -1\end{array}\right).
\end{equation}
The matrix $u_0=1$. Inequality (\ref{CC8}) is saturated since
\begin{equation}\label{CC14}
S_{\rm vN}=0,\qquad S(F)=\ln 2
\end{equation} and
\begin{equation}\label{CC15}
S_{\rm vN}+ S(F)=\ln 2\geq \ln 2.
\end{equation}
Also the inequality for R\'{e}nyi entropy is saturated
\begin{equation}\label{CC16}
R_{\alpha}(u_0)+R_{\beta}(Fu_0)=\ln 2\geq \ln 2.
\end{equation}
In the considered example, the $\mid q\rangle$ and $\mid p\rangle$ vectors
form that is called ``mutually unbiased bases'' \cite{planat,Klimov,paz,Rosu}.

One should note that there are Shannon entropic uncertainty relations for
distributions associated with set of mutually unbiased bases \cite{Balles}
and with pairs of orthogonal bases \cite{uff}. In the case where mutually unbiased bases
 are connected by quantum Fourier transform, our result (\ref{CC7a}) coincides with \cite{uff}.

In \cite{Rui-JRLR} group average Shannon and R\'{e}nyi entropies were introduced.

Due to invariance of Haar measure, one can conclude that the group average
Shannon tomographic entropy satisfies the inequality
\begin{equation}\label{CC17}
\bar{H}=\int H(u)\,du\geq \frac{1}{2}\ln N.
\end{equation}
Also for group average R\'{e}nyi entropy (\ref{13}), one has
\begin{equation}\label{CC18}
\bar{R}_{\alpha\beta}=\int R_{\alpha}(u)\,du+\int R_{\beta}(u)\,du \geq \ln N,
\qquad \frac{1}{\alpha}+\frac{1}{\beta}=2.\end{equation}

To illustrate the inequalities obtained, let us now discuss the mixed state of qubit with
diagonal density matrix
\begin{equation}\label{CC19}
\rho=\left(\begin{array}{cc}
a & 0\\
0 & b\end{array}\right), \qquad a+b=1.
\end{equation}
Then inequality (\ref{CC6}) can be visualized as follows.

Von Neumann entropy of this state
\begin{equation}\label{CC20}
S_{\rm vN}=-a\ln a-b\ln b.
\end{equation}
The density matrix subjected by quantum Fourier transform (\ref{CC13}) reads
\begin{equation}\label{CC21}
F^\dagger\rho F=\left(\begin{array}{cc}
1/2 & (a-b)/2\\
(a-b)/2 & 1/2\end{array}\right).
\end{equation}
Its tomographic entropy
\begin{equation}\label{CC22}
H(Fu_o)=\ln 2,\qquad u_0=1.
\end{equation}
Thus inequality (\ref{CC6}) looks as follows:
\begin{equation}\label{CC23}
-a\ln a-b\ln b+\ln 2\geq\ln 2,
\end{equation}
which only means that von Neumann entropy is nonnegative. But inequality (\ref{CC6a})
gives better estimation since the number $\ln 2$ is replaced by a smaller number. In fact,
the tomographic-probability vector of the qubit state
\begin{equation}\label{CC24}
\vec w=\left(\begin{array}{c}
a\\b\end{array}\right)
\end{equation}
is associated to the probability-amplitude vector with positive components
\begin{equation}\label{CC24a}
\vec W=\left(\begin{array}{c}
\sqrt a\\
\sqrt b\end{array}\right)
\end{equation}
Then after making the quantum Fourier transform of this vector, we get the column vector
\begin{equation}\label{CC25}
\vec W_F=\frac{1}{\sqrt 2}\left(\begin{array}{c}
\sqrt a+\sqrt b\\
\sqrt a-\sqrt b\end{array}\right).
\end{equation}
The probability-distribution vector associated to the above  probability-amplitude vector
reads
\begin{equation}\label{CC26}
\vec w_F=\left(\begin{array}{c}
(1/2)+{\sqrt {ab}}\\
(1/2)-{\sqrt {ab}}\end{array}\right).
\end{equation}
Thus we apply inequality relating Shannon entropies to two vectors (\ref{CC24a})
and (\ref{CC26}) and obtain
\begin{equation}\label{CC27}
-a\ln a-b\ln b-\left(\frac{1}{2}+{\sqrt{ab}}\right)\ln\left(\frac{1}{2}+{\sqrt{ab}}\right)
-\left(\frac{1}{2}-{\sqrt{ab}}\right)\ln\left(\frac{1}{2}-{\sqrt{ab}}\right)\geq\ln 2,
\end{equation}
or
\begin{equation}\label{CC28}
S_{\rm vN}-\left(\frac{1}{2}+{\sqrt{ab}}\right)\ln\left(\frac{1}{2}+{\sqrt{ab}}\right)
-\left(\frac{1}{2}-{\sqrt{ab}}\right)\ln\left(\frac{1}{2}-{\sqrt{ab}}\right)\geq\ln 2.
\end{equation}
This inequality is not that obvious though we know that $S_{\rm vN}\geq 0$.

Some inequalities for unitary matrix can be obtained.

Let us consider $N$$\times$$N$-unitary matrix $u_{jk}$. One has the inequality
\begin{equation}\label{AAA}
-\sum_{j=1}^N\left(|u_{jk}|^2\ln|u_{jk}|^2+|(Fu)_{jk}|^2\ln|(Fu)_{jk}|^2\right)\geq\ln N
\end{equation}
or
\begin{equation}\label{BBB}
-\sum_{j=1}^N\sum_{k=1}^N\left(|u_{jk}|^2\ln|u_{jk}|^2+|(Fu)_{jk}|^2\ln|(Fu)_{jk}|^2\right)
\geq N\ln N,
\end{equation}
where $F_{jk}$ is the Fourier transform matrix. Integrating inequality (\ref{AAA}) over
the unitary group with Haar measure normalized as in (\ref{3})
one has the inequality
\begin{equation}\label{DDD}
-\int\left(\sum_{j=1}^N|u_{jk}|^2\ln|u_{jk}|^2\right)du\geq\frac12\ln N.
\end{equation}

We demonstrated on the example of qubit that for tomograms of the spin states connected
by quantum Fourier transforms one has constraints in the form of inequalities for Shannon
tomographic entropies. One can demonstrate analogous constraints for R\'{e}nyi tomographic
entropies too.

\section{Conclusions}

We point out that there exist several inequalities for Shannon and R\'{e}nyi entropies
associated to spin quantum state tomograms. These inequalities provide extra information
theory constraints in addition to known subadditivity and strong subadditivity conditions.
Physical and information meaning of the inequalities obtained needs extra clarification.

\section*{Acknowledgments}

The study was supported by the Russian Foundation for Basic Research under Project
No.~07-02-00598.
The authors thank the University of Lisbon where this work was initiated and partially
done for hospitality. M.A.M. thanks the Organizers of the International Workshop
``Nonlinear Physics. Theory and Experiment. V'' (Gallipoli, Lecce,
Italy, 2008) for kind hospitality and the Russian Foundation for
Basic Research for Travel Grant No.~08-02-08174.

\end{document}